\documentclass[fleqn,12pt,twoside]{article}
\usepackage{espcrc1}

\usepackage{graphicx}
\usepackage{epsfig}
\usepackage[figuresright]{rotating}

\newcommand{\AmS}{{\protect\the\textfont2
  A\kern-.1667em\lower.5ex\hbox{M}\kern-.125emS}}

\title{Models of the Solar Wind Interaction with Local Interstellar Cloud}

\author{V.~V. Izmodenov\address{Division of Aeromechanics and Gas Dynamics,
        Department of Mechanics and Mathematics, Lomonosov Moscow State University,
        Moscow, 119899, Russia}%
      }

\begin{document}
\maketitle

\begin{abstract}

This paper reviews the theoretical approaches and existing models of the solar wind interaction
with the Local Interstellar Cloud (LIC).
Models discussed take into account the multi-component nature of the solar
wind and local interstellar medium.
Basic results of the modeling and their possible applications
to interpretation of space experiments are summarized.
Open questions of global modeling of the solar wind/LIC interaction  and future perspectives
are discussed.
\end{abstract}

\section{INTRODUCTION}

The Local Interstellar Cloud (LIC) is a cloud of partly ionized
plasma surrounding the Solar System. The plasma component of LIC
interacts with the solar wind plasma and forms the heliospheric
interface (Figure 1). The heliospheric interface is a complex
structure, where the solar wind and interstellar plasma,
interplanetary and interstellar magnetic fields, interstellar
atoms of hydrogen, galactic and anomalous cosmic rays (GCRs and
ACRs) and pickup ions play prominent roles.

Although a space mission into the Local Interstellar Cloud is
becoming now more realisable, there are no yet direct observations
inside the heliospheric interface. Therefore, at the present time
the heliospheric interface structure and local interstellar
parameters can be derived only from remote experiments and
measurements. Currently, backscattered solar Ly-$\alpha$
radiation, pickup ions, anomalous cosmic rays, and Voyager
measurements of distant solar wind are the major sources of
information on the heliospheric interface structure and position
of the termination shock \cite{issibook_96}. kHz emission detected
by Voyager can put some constraints. Recently, it was shown that
study of Ly-$\alpha$ absorptions toward nearby stars can serve as
remote diagnostics of the heliospheric interface and, in
particular, the hydrogen wall around the heliopause (e.g.,
\cite{Linsky_Wood96} - \cite{iwl2002}). In the foreseeable future,
remote diagnostics will be also possible with images of
heliospheric energetic neutrals (ENAs) \cite{Gruntman_2001}.
 To reconstruct the structure of the interface and physical processes inside
 the interface on the basis of remote observations, a theoretical
model should be employed.

\begin{figure}[htb] \label{fig1}

\psfig{figure=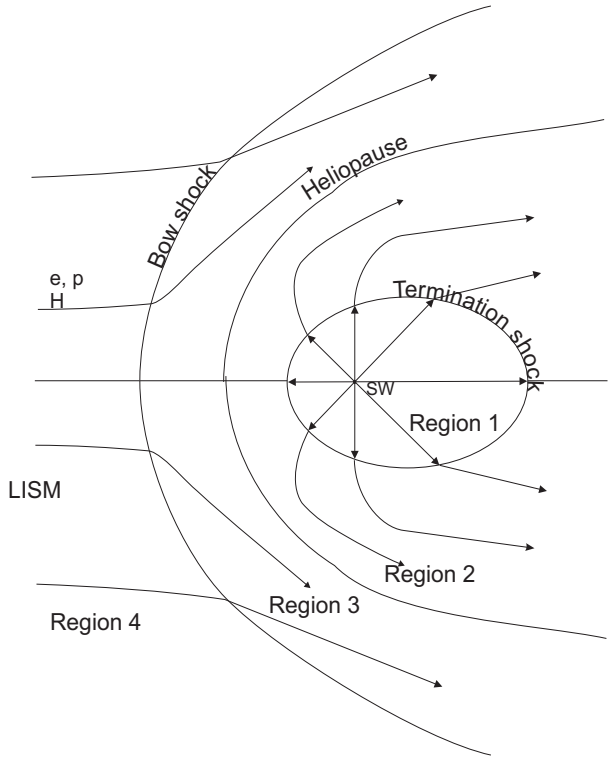,height=10cm,width=8cm,angle=0}
\caption{ The heliospheric interface is the region of the solar
wind interaction with LIC. The heliopause is a contact
discontinuity, which separates the plasma wind from interstellar
plasmas. The termination shock decelerates the supersonic solar
wind. The bow shock may also exist in the interstellar medium. The
heliospheric interface can be divided into four regions with
significantly different plasma properties: 1) supersonic solar
wind; 2) subsonic solar wind in the region between the heliopause
and termination shock; 3) disturbed interstellar plasma region (or
"pile-up" region) around the heliopause; 4) undisturbed
interstellar medium. }

\end{figure}

Theoretical studies of the heliospheric interface were performed
over more than four decades after pioneering papers by Parker
\cite{parker61} and Baranov et al. \cite{baranov71}. However, a
complete theoretical model of the heliospheric interface has not
been constructed yet. The difficulty in doing this is connected
with the multi-component nature of both the LIC and the solar
wind. The LIC consists of at least five components: plasma
(electrons and protons), hydrogen atoms, interstellar magnetic
field, galactic cosmic rays, and interstellar dust. The
heliospheric plasma consists of original solar particles (protons,
electrons, alpha particles, etc.), pickup ions and the anomalous
cosmic ray component. The pickup ion component is a result of
ionization of those interstellar H atoms that penetrate into the
heliosphere through the heliospheric interface. A part of the
pickup ions is accelerated to high energies of ACRs. ACRs may also
modify the plasma flow upstream of the termination shock and in
the heliosheath. Spectra of ACRs can serve as remote diagnostics
of the termination shock. For a recent review on ACRs see
\cite{Fichtner_2001}.

To construct a theoretical model of the heliospheric interface,
one needs to choose a specific approach for each interstellar and
solar wind component. Interstellar and solar wind protons and
electrons can probably be described as fluids. At the same time
interstellar H atom flow requires kinetic description. For pickup
ion and cosmic ray components, the kinetic approach is also
required. However, for interpretations that are not directly
connected to pickup ions and ACRs, a cruder model can be used.

\begin{table}[hbt]
  \centering
  \caption{Number Densities and Pressures of Solar Wind Components}\label{tab:sw_pressures}
  \begin{tabular*}{\textwidth}{@{}l@{\extracolsep{\fill}}ccccc}
\hline
Component     &  4-5 AU & & 80 AU & \\
      & Number Density & Pressure & Number Density & Pressure \\
               & cm$^{-3}$ & eV/cm$^{-3}$    & cm$^{-3}$ & eV/cm$^{-3}$ \\ \hline
Original solar & 0.2-0.4   & 2.-4. (thermal) & ($7-14$) $\cdot$
10$^{-4}$ & 10$^{-3}$ - 10$^{-4}$ \\
wind protons &    & $\sim 200$ (dynamic) & & $\sim 0.5 -1. $ (dynamic) \\
Pickup ions & $5.1 \cdot 10^{-4}$ & 0.5 & $\sim 2 \cdot 10^{-4}$ & $\sim 0.15$ \\
Anomalous  &  & & &   \\
cosmic rays &  & & & 0.01 - 0.1 \\
\hline
  \end{tabular*}
\end{table}

This paper focuses on the models of the global heliospheric
interface structure. Under global models I understand those models
that study the whole interaction region, including the termination
shock, the heliopause and possible bow shock. In this sense, this
paper should not be considered as a complete review of progress in
the field. Many different approaches were used to look into
different aspects of the solar wind interaction with LIC
connecting with pickup ion transport and acceleration, with the
termination shock structure under influence of ACRs and pickup
ions. For more complete overview see recent reviews
\cite{Zank_1999},\cite{Fichtner_2001}.

The structure of the paper is the following: The next section
briefly describes our current knowledge of the local interstellar
and solar wind parameters. Section 3 discusses theoretical
approaches to be used for the interstellar and solar wind
components. Section 4 gives an overview of heliospheric interface
models. Section 5 describes basic results of the Baranov-Malama
model of the heliospheric interface and its future developments.
In section 6 we demonstrate possible analyses of space experiments
on the basis of a theoretical model of the heliospheric interface.
Section 7 underlines current problems in the modeling of the
global heliosphere and discusses future perspectives.

\section{BRIEF SUMMARY OF OBSERVATIONAL KNOWLEDGE}

Choice of an adequate theoretical model of the heliospheric
interface depends on boundary conditions, i.e. on undisturbed
solar wind and interstellar parameters.

\subsection{Solar wind observations}
At the Earth's orbit the flux of interstellar atoms is quite
small, and the solar wind can be considered undisturbed.
Measurements of pickup ions and ACRs also show that these
components do not have dynamical influences on the original solar
wind particles at the Earth's orbit. Therefore, solar wind
parameters at the Earth's orbit can be taken as inner boundary
conditions.

It has been shown by many authors that pickup and ACR components
dynamically influence the solar wind at large heliocentric
distances. Observable evidence of such influence is, for example,
deceleration of the solar wind detected by Voyager
\cite{Richardson_2001}. Table~\ref{tab:sw_pressures} presents
estimates of dynamic importance of the heliospheric plasma
components at small and large heliocentric distances. The table
shows that pickup ion thermal pressure can be up to 30-50 \% of
the dynamic pressure of solar wind.

\begin{table}[hbt]
\newlength{\digitwidth} \settowidth{\digitwidth}{\rm 0}
\catcode`?=\active \def?{\kern\digitwidth}
\caption{Local Interstellar Parameters}
\label{tab:interstellar_parameters}
\begin{tabular*}{\textwidth}{@{}l@{\extracolsep{\fill}}llll}
\hline

Parameter                 & Direct measurements/estimations \\
\hline
Sun/LIC relative velocity      &  25.3 $\pm$ 0.4 km s$^{-1}$ (direct He atoms $^{1}$) \\
                               & 25.7 km s$^{-1}$ (Doppler-shifted \\
                               &  absorption lines $^{2}$) \\
Local interstellar temperature & 7000 $\pm$ 600 K (direct He atoms $^{1}$) \\
                               & 6700 K (absorption lines $^{2}$) \\
LIC H atoms number density     & 0.2 $\pm$ 0.05 cm$^{-3}$ (estimate based on \\
                               & pickup ion observations $^{3}$) \\
LIC proton number density      & 0.03 - 0.1 cm$^{-3}$ (estimate based on  \\
                               &pickup ion observations $^{3}$) \\
Local Interstellar magnetic field &  Magnitude: 2-4 $\mu$G \\
                                  &  Direction: unknown   \\
Pressure of low energetic part of cosmic rays &  $\sim$0.2 eV
cm$^{-3}$ \\
\hline \multicolumn{5}{@{}p{120mm}}{$^{1}$\cite{Witte_1996};
$^{2}$ \cite{Lallement_1996}; $^{3}$ \cite{Gloeckler_1997}}
\end{tabular*}
\end{table}

\subsection{Interstellar parameters}

Local interstellar temperature and velocity can be inferred from
direct measurements of interstellar atoms of helium by Ulysses/GAS
instrument \cite{Witte_1996}. Atoms of interstellar helium
penetrate the heliospheric interface undisturbed, because of the
small strength of their coupling with interstellar and solar wind
protons. Indeed, due to small cross sections of elastic collisions
and charge exchange with protons, the mean free path of these
atoms is larger than the heliospheric interface. Independently,
the velocity and temperature in the Local Interstellar Cloud can
be deduced from analysis of absorption features in the stellar
spectra \cite{Lallement_1996}. However, this method provides mean
values along the line of sight in the LIC. A comparison of local
interstellar temperatures and velocities derived from stellar
absorption with those derived from direct measurements of
interstellar helium shows quite good agreement (see Table 2).

Other local parameters of the interstellar medium, such as
interstellar H atom and electron number densities, and strength
and direction of the interstellar magnetic field, are not well
known. In the models they can be considered as free parameters.
However, measurements of interstellar H atoms and their
derivatives as pickup ions and ACRs provide important constraints
on local interstellar densities and total pressure. The neutral H
density in the inner heliosphere depends on filtration the neutral
H atoms in the heliospheric interface due to charge exchange.
Since interstellar He is not perturbed in the interface, local
interstellar number density of H atoms can be estimated from the
neutral hydrogen to the neutral helium ratio in the LIC,
$R(HI/HeI)_{LIC}$: $ n_{LIC} (HI) = R(HI/HeI)_{LIC} n_{LIC}(HeI)
$. The neutral He number density in the heliosphere has been
recently determined to be very likely around $0.013- 0.018$
cm$^{-3}$ (\cite{Witte_1996} - \cite{Moebius_1996}). Interstellar
ratio HI/HeI is likely in the range of 10-14. Therefore, expected
interstellar H atom number densities are in the range of  $0.13 -
0.25$ cm$^{-3}$. It was shown by modeling
\cite{Baranov_Malama_1995}, \cite{Izmodenov_1999} that the
filtration factor, which is the ratio of neutral H density inside
and outside the heliosphere, is a function of interstellar plasma
number density. Therefore, the number density of interstellar
protons (electrons) can be estimated from this filtration factor
\cite{Lallement_1996}. Independently, the electron number density
in the LIC can be estimated from abundances ratios of ions of
different ionization states \cite{Lallement_1996}.

Note that there are other methods to estimate interstellar H atom
density inside the heliosphere, based on their influence on the
distant solar wind \cite{Richardson_2001} or from ACR spectra
\cite{Gurnett_Kurth}. Recent estimates of the location of the
heliospheric termination shock using transient decreases of cosmic
rays observed by Voyager 1 and 2 also provide constraints on the
local interstellar parameters \cite{Webber_2001}. However,
simultaneous analysis of different types of observational
constraints has not been done yet. Theoretical models should be
employed to make such analysis.
Table~\ref{tab:interstellar_parameters} presents a summary of our
knowledge of local interstellar parameters. Using these
parameters, we estimate local pressures of different interstellar
components (Table 3). All pressures have the same order of
magnitude.
 This means that theoretical models should not neglect any of these interstellar components.
Dynamical pressure of interstellar H atoms is larger than all other pressures.
A part of H atoms, ACRs and GCRs penetrate into the heliosphere, which makes
their real dynamical influence on the heliospheric plasma interface
difficult to estimate.

\begin{table}[hbt]
  \centering
  \caption{Local Pressures of Interstellar Components}\label{tab: lic_pressures}
  \begin{tabular*}{\textwidth}{@{}l@{\extracolsep{\fill}}ll}
\hline
Component     &  Pressure estimation, dyn cm$^{-2}$ \\
\hline
{\bf Interstellar plasma component}  &   \\
Thermal pressure               & $(0.6 - 2.0) \cdot 10^{-13}$ \\
Dynamic pressure               & $(1.5 - 6) \cdot 10^{-13}$  \\
\hline
{\bf H atoms}                & \\
Thermal pressure               &$(0.6 - 2.0) \cdot 10^{-13}$ \\
Dynamic pressure               &$(4.0 - 9.0) \cdot 10^{-13}$  \\
\hline
Interstellar magnetic field   &$(1.0 - 5.0) \cdot 10^{-13}$  \\
\hline
Low energy part of GCR         &$(1.0 - 5.0) \cdot 10^{-13}$  \\
\hline
  \end{tabular*}
\end{table}

\section{OVERVIEW OF THEORETICAL APPROACHES}

In this section we consider theoretical approaches for components
involved in the dynamical processes in the heliospheric interface.

Generally, any gas can be described on a kinetic or a hydrodynamic
level. In the kinetic approach, macroscopic parameters of a gas of
$s$-particles (or, briefly, $s$-gas) can be expressed through
integrals of velocity distribution function $f_s(\vec{r}, \vec{w},
t)$: $n_{s} = \int f_{s} d \vec{w}$, $ \vec{V}_{s} = (\int \vec{w}
f_s d \vec{w})/n_s $, $P_{s, ij}= m_{s} \int (w_i -V_{s,i})(w_j -
V_{s,j}) f_s d \vec{w}$, $\vec{q}_s = 0.5 m_s \int
(\vec{w}-\vec{V_s})^2 (\vec{w} - \vec{V}_s) f_s d \vec{w}$, where
$n_s$ is the number density of $s$-gas, $\vec{V}_s$ is the bulk
velocity of $s$-gas, $P_{s, ij}$ are components of the stress
tensor $\widehat{P}_s$, $\vec{q}_s$ is the thermal flux vector,
$m_s$ is the mass of individual $s$-particle. In the hydrodynamic
approach, some assumptions should be made to specify the stress
tensor $\widehat{P}_s$, and the thermal flux vector, $\vec{q}_s$
to make hydrodynamic system closed. For example, these values can
be calculated by the Chapman-Enskog method, assuming $Kn = l/L <<
1$, where $l$ and $L$ are the mean free path of the particles and
characteristic size of the problem, respectively. The zero
approximation of the Chapman-Enskog method gives local Maxwellian
distribution, and the gas can be considered as an ideal gas, where
the stress tensor reduces to scalar pressure $P$ and $\vec{q}=0$.

\subsection{H atoms}

Interstellar atoms of hydrogen form the most abundant component in
the circumsolar local interstellar medium (see, Table 2). These
atoms penetrate deep into the heliosphere and interact with
interstellar and solar wind plasma protons. The cross sections of
elastic H-H, H-p collisions are negligible as compared with the
charge exchange cross section \cite{izmod_et_al_2000}. Charge
exchange with solar wind/interstellar protons determines the
properties of the H atom gas in the interface.
 Atoms, newly created by charge exchange, have the local properties of protons.
Since plasma properties are different in the four regions of the
heliospheric interface shown in Figure 1, the H atoms can be
separated into four populations, each having significantly
different properties. The strength of H atom-proton coupling can
be estimated through the calculation of mean free path of H atoms
in plasma. Generally, the mean free path (with respect to the
momentum transfer) of s-particle in t-gas can be calculated by the
formula: $l= m_s w^2_s / ( \delta M_{st}/\delta t)$. Here, $w_s$
is the individual velocity of s-particle, and $\delta M_{st} /
\delta t$ is individual s-particle momentum transfer rate in
t-gas.

Table~\ref{tab:mfps} shows the mean free paths of H atoms with
respect to charge exchange with protons. The mean free paths are
calculated for typical atoms of different populations at different
regions of the interface in the upwind direction. For every
population of H atoms, there is at least one region in the
interface where the Knudsen number $Kn \approx 0.5-1.0$.
Therefore, the kinetic Boltzmann approach must be used to describe
interstellar atoms in the heliospheric interface.

\begin{table}[hbt]
\catcode`?=\active \def?{\kern\digitwidth}
\caption{Mean free paths of H-atoms in the heliospheric interface
with respect to charge exchange with protons, in AU.}
\label{tab:mfps}
\begin{tabular*}{\textwidth}{lrrrr}
\hline
Population & At TS & At HP & Between HP and BS & LISM         \\
\hline
4 (primary interstellar)    & $ 150$ & $100$ & $110$ & $870$ \\
3 (secondary interstellar)  & $ 66$ & $40$ &  $58$   & $190$ \\
2 (atoms originating in the heliosheath) & $830$ & $200$ & $110$ & $200$ \\
1 (neutralized solar wind)  & $ 16000$ & $510$ & $240$ & $490$ \\
\hline
\end{tabular*}
\end{table}

The velocity distribution of H atoms $f_{\rm H}(\vec{r},
\vec{w}_{\rm H}, t)$ may be calculated from the linear kinetic equation introduced in \cite{bm93}:
\begin{eqnarray}
\label{eqBoltz}
\frac{\partial f_{\rm H}}{\partial t}+ \vec{w}_{\rm H} \cdot
\frac{\partial f_{\rm H}}{\partial \vec{r} }
+ \frac{\vec{F}}{m_{\rm H}} \cdot
\frac{\partial f_{\rm H}}{\partial
\vec{w}_{\rm H} } =
- f_{\rm H}
\int | \vec{w}_{\rm H} - \vec{w}_p |
\sigma^{\rm HP}_{ex} f_p
(\vec{r}, \vec{w}_p) d \vec{w}_p \\ \nonumber
+ f_p (\vec{r}, \vec{w}_{\rm H}) \int | \vec{w}_{\rm H}^* - \vec{w}_{\rm H} |
\sigma^{\rm HP}_{ex} f_{\rm H}
(\vec{r}, \vec{w}_{\rm H}^* ) d \vec{w}_{\rm H}^*
- ( \nu_{ph} + \nu_{\rm impact} ) f_{\rm H} ( \vec{r}, \vec{w}_{\rm H} ).
\end{eqnarray}
Here $ f_{\rm H} ( \vec{r}, \vec{w}_{\rm H}) $ is the distribution
function of H atoms; $ f_p( \vec{r}, \vec{w}_p ) $ is the local
distribution function of protons; $ \vec{w}_p $ and $ \vec{w}_{\rm
H}$ are the individual proton and H atom velocities, respectively;
$ \sigma^{\rm HP}_{ex} $ is the charge exchange cross section of
an H atom with a proton; $ \nu_{ph} $ is the photoionization rate;
$ m_{\rm H}$ is the atomic mass; $ \nu_{\rm impact} $ is the
electron impact ionization rate; and $ \vec{F} $ is the sum of the
solar gravitational force and the solar radiation pressure force.
The plasma and neutral components interact mainly by charge
exchange. However, photoionization, solar gravitation, and
radiation pressure, which are taken into account in equation
($\ref{eqBoltz}$), are important at small heliocentric distances.
Electron impact ionization may be important in the heliosheath
(region 2). The interaction of the plasma and H atom components
leads to the mutual exchanges of mass, momentum and energy. These
exchanges should be taken into account in the plasma equations
through source terms, which are integrals of $f_H( \vec{r},
\vec{w},t)$.

\subsection{Solar wind and interstellar electron and proton components}

Basic assumptions necessary to employ a hydrodynamic approach for
space plasmas were reviewed in \cite{baranov2000}. In particular,
it was concluded in the paper that interstellar and solar wind
plasmas can be treated hydrodynamically. Indeed, the mean free
path of the charged particles in the local interstellar plasma is
less than 1 AU, which is much smaller than the size of the
heliospheric interface itself. Therefore, the local interstellar
plasma is collisional plasma, and a hydrodynamic approach can be
used to describe it. Solar wind plasma is collisionless, because
the mean free path of the solar wind particles is much larger than
the size of the heliopause. Therefore, the heliospheric
termination shock (TS) is a collisionless shock. A hydrodynamic
approach can be justified for collisionless plasmas when
scattering of charged particles on plasma fluctuations is
efficient ("collective plasma processes"). In this case, the mean
free path $l$ with respect to collisions is replaced by
$l_{coll}$, the mean free path of collective processes, which is
assumed to be less than the characteristic length of the problem
L:  $ l_{coll} << L$. However, the integral of "collective
collisions" is too complicated to be used to calculate the
transport coefficient for collisionless plasmas.

One-fluid description of heliospheric and interstellar plasmas is
commonly used in the global models of the heliospheric interface.
However, since measurements of the solar wind show different
electron and proton temperatures, two-fluid approach is more
appropriate. The temperatures may remain different up to the
termination shock and beyond due to the weak energy exchange
between protons and electrons.

Hydrodynamic Euler equations for proton and electron components,
which take into account the influence of other components such as
interstellar H atoms, pickup ions, cosmic rays, electric and
magnetic fields, are written below. Mass balance or continuity
equations are

\begin{eqnarray}\label{density}
\frac{\partial n_s}{\partial t} + \nabla \cdot (n_s
\vec{V}_s) = q_{1, s}, (s=e,p)
\end{eqnarray}

Index $e$ denotes electrons, index $p$ denotes solar wind protons.
 $q_{1,e} = n_H \cdot (\nu_{ph}+ \nu_{impact})$, $q_{1,p} =  -
\int u \sigma^{\rm HP}_{ex}(u) f_p(\vec{w}) f_H(\vec{w}_H) d
\vec{w} d \vec{w}_H$ are sources and sinks due to charge exchange,
photoionization and electron impact ionization. Here, $u =
|\vec{w}_H - \vec{w}| $ is the relative atom-proton velocity, and
$\vec{w}_H$ and $\vec{w}$ are individual velocities of H atoms and
protons, respectively. Momentum balance equations are
\begin{eqnarray}\label{momentum}
\frac{\partial (n_s m_s \vec{V}_s) }{\partial t} + \nabla P_s +
m_s \nabla \cdot (n_s \vec{V}_s \otimes \vec{V}_s) - n_s e_s (\vec{E}
+\frac{1}{c}[\vec{V}_s \times \vec{B}] ) + \sum_r \vec{R}_{s r} =
m_s \vec{q}_{2, s}
\end{eqnarray}

where $ \vec{q}_{2,e} = \int (\nu_{ph}+ \nu_{impact}) \vec{w}_H
f_H(\vec{w}_H) d \vec{w}_H $, $\vec{q}_{2,p} = - \int \int u
\sigma^{\rm HP}_{ex}(u) \vec{w}_p f_H(\vec{w}_H) f_p(\vec{w}_p) d
\vec{w}_H d \vec{w}_p $ and $\vec{R}_{s r} $ is the rate of
momentum transfer from the $r$-gas component to the $s$-gas
component. The symbol $\otimes$ represents the dyadic product. The
momentum transfer term $\vec{R}_{sr}$ can be expressed in a
general form through the collision integral $S_{s r}$ of kinetic
equation of the $s$-gas component: $ \vec{R}_{sr} = - \int m_s
\vec{c}_s S_{sr} d c_s$, where $ \vec{c}_s = \vec{w}_s -
\vec{V}_s$. That $ \vec{R}_{sr} + \vec{R}_{rs} = 0$ is a
consequence of this definition of $\vec{R}_{sr}$.

Heat balance equations have the following form:
\begin{eqnarray}\label{energy}
\frac{\partial}{\partial t} \left( \frac{3}{2} P_s \right) +
\nabla  \left( \frac{3}{2}P_s \vec{V}_s \right) + P_s \nabla \cdot
\vec{V}_s = \sum_r Q_{sr} + m_s q_{3,s} - m_s \vec{q}_{2,s}
\cdot \vec{V}_s
\end{eqnarray}
with $ q_{3,e}= \int{ \left( \nu_{ph} + \nu_{impact} \right)
\frac{\vec{w}_H^2}{2} f_H(\vec{w}_H) d \vec{w}_H }$,

$ q_{3,p}= - \int \int u \sigma^{\rm HP}_{ex}(u)
\frac{\vec{w}_p^2}{2} f_H(\vec{w}_H) f_p(\vec{w}_p) d \vec{w}_p d
\vec{v}_H $. $ Q_{s r}$ is the heat source due to interactions
between between particles of $s$ and $r$ components.
  Those terms can be expressed in a general form through the collision term $S_{s r}$ of the kinetic equation $
Q_{s r} = - \int \frac{m_s c^2_s}{2} S_{p r}
d c_s$.
As a result of this interaction, we have a relation connecting $Q_{sr}$ and $Q_{rs}$: $Q_{s r}+ Q_{r s}
= -\vec{R}_{s r} \cdot (\vec{V}_{s} -
\vec{V}_{r} ) (\forall s \neq r$).

System (\ref{density})-(\ref{energy}) should be added by the state equations:
$P_\alpha = n_\alpha k T_\alpha$ ($\alpha = e,p)$, where $k$ is Boltzman constant, and Maxwell equations:
\begin{equation}\label{Maxwell}
\nabla \times \vec{E}= - \frac{1}{c} \frac{ \partial
\vec{B}}{\partial t}; \nabla \cdot \vec{E} = 4 \pi \rho_e;
\nabla \times \vec{B} = \frac{4 \pi}{c} \vec{j}; \nabla \cdot \vec{B} = 0
\end{equation}
where $\rho_e $ is the charge density, $e$ is the charge of
electron, and $\vec{j}$ is current density. The displacement current has been dropped. Note, that charge and
current densities of all charged populations should be taken into
account in (5). Neglecting cosmic ray charges and currents we have
$\rho_e = e (n_p+n_{pui}-n_e)$ and $\vec{j} = e (n_p \vec{V}_p -
n_e \vec{V}_e + n_{pui} \vec{V}_{pui}) $. The number density of
pickup ions, $n_{pui}$, and the bulk velocity of pickup ions,
$\vec{V}_{pui}$ are integrals of pickup proton velocity
distribution function: $n_{pui} = \int f_{pui}(\vec{w}) d
\vec{w}$, $ \vec{V}_{pui} = (\int \vec{w} f_{pui}(\vec{w}) d
\vec{w})/n_{pui} $.

Note that in equations (\ref{density})-(\ref{energy}) we assume
that pickup electrons are indistinguishable from original solar
wind electrons, while pickup protons are considered as a separate
population.

The expressions for various interaction terms $\vec{R}_{sr}$, $Q_s$ ($ s = e,p$)must be specified.
Electron-proton collision terms can be taken in the form given by Braginski
\cite{braginski}:
\begin{equation}\label{R}
\vec{R}_{ep} = -\frac{m_e n_e}{\tau_e} \vec{u}_e
\end{equation}
\begin{equation}\label{Q}
Q_{pe} = Q_{ep}- R_{ep} \vec{u}_e = \frac{3
n_e}{\tau_e} \frac{m_e}{m_p} k (T_e - T_p)
\end{equation}
Here, $n_e$ is the electron number density; $T_e$ and $T_p$ are
electron and proton densities, respectively; $m_e$ and $m_p$ are
the electron and proton masses; $\vec{u}_e$ is the electron
velocity relative to the proton rest frame. Parameter $\tau_e$
characterizes the coupling between electrons and protons and
corresponds to the electron collision time for collisional plasma
\cite{braginski}. In collisionless heliospheric plasma, additional
assumptions are needed to determine  $\tau_e$; otherwise, it can
be considered as free parameter.

\subsection{Pickup ions}

To study pickup ion dynamical influence on the distant solar wind,
the termination shock structure and, finally, on the global
heliospheric interface structure, details of the process of
charged particle assimilation into the magnetized plasma are
needed. A newly created ion under the influence of the steady
solar wind electric and magnetic fields executes a cycloidal
trajectory with the guiding center, which is drifting at the bulk
velocity of the solar wind. Assuming that the gyroradius is much
smaller than the typical scale length, one can average velocity
distribution function over the gyratory motion. Initial ring-beam
distribution of pickup ions is unstable. Basic processes that
determine evolution of pickup ion distribution are pitch-angle
scattering, energy diffusion in the wave field  generated by both
pickup ions and the solar wind waves, convection, adiabatic
cooling in the expanding solar wind, and injection of newly
ionized particles. The most general form of the relevant transport
equation to describe the evolution of gyrotropic velocity
distribution function $f_{pui} = f_{pui}(t, \vec{r}, v, \mu)$ of
pickup ions in a background plasma moving at a velocity
$\vec{V}_{sw}$ were written in \cite{isenberg97},
\cite{chalov_fahr98}. $f_{pui}$ is a function of the modulus of
velocity in the solar wind rest frame, and $\mu$ is the cosine of
pitch angle.

Complete assimilation of pickup ions into the solar wind would
result in a great increase in the temperature with increasing
heliocentric distance, which is not observed. Therefore, the solar
wind and pickup protons represent two distinct proton populations.
Nevertheless, the radial temperature profile of protons measured
by Voyager 2 shows a smaller decrease as compared with the
adiabatic cooling. A fraction of heating of solar wind protons may
be connected with pickup generated waves \cite{Williams_1995}.
Many aspects of pickup ion evolution were studied ( e.g.,
\cite{chalov_fahr98}; for review, see \cite{Zank_1999},
\cite{Fichtner_2001}). However, today it still seems to be
impossible to take into account all details of the assimilation
process of pickup ions into the solar wind in the global models of
the heliospheric interface structure. Instead, one may try to use
the hydrodynamic approach. In this approach, equations
(\ref{density})-(\ref{energy}) written for pickup ions represent
the balance of their mass, momentum and energy. The right sides of
the equations include sources of pickup ions due to ionization
processes:
\[
q_{1,pui} = n_H \nu_{ph}+ \int u \sigma^{\rm HP}_{ex}(u)
f_H(\vec{w}_H) f_p (\vec{w}) d \vec{w} d \vec{v}_H
\]
\[
\vec{q}_{2,pui}= \int (\nu_{ph}+\nu_{impact}) \vec{v}_H
f_H(\vec{v}_H) d \vec{v}_H + \int \int
 u \sigma^{\rm HP}_{ex}(u) \vec{w}_H f_H(\vec{w}_H) f_p(\vec{w}_p) d \vec{w}_H d
\vec{w}_p +
\]
\[
\int \int u \sigma^{\rm HP}_{ex}(u) (\vec{w}_H - \vec{w}_i)
f_H(\vec{w}_H) f_{pui}(\vec{w}_i) d \vec{w}_H d \vec{w}_i
\]
\[
q_{3,pui}= \int \left( \nu_{ph} + \nu_{impact} \right)
\frac{\vec{w}_H^2}{2} f_H(\vec{w}_H) f_p(\vec{w}_p) d \vec{w}_p d
\vec{w}_H
\]
\[
 +
\int \int  u \sigma^{\rm HP}_{ex}(u) \frac{\vec{w}_H^2-
\vec{w}_i^2}{2} f_H(\vec{w}_H) f_{pui}(\vec{w}_i) d \vec{w}_i d
\vec{w}_H \nonumber
\]
To complete the model, one should also specify interaction terms
$\vec{R}_{pui, r}$, $Q_{pui, r}$ ($r \neq pui$). The specification
of these terms for pickup ion-proton interactions requires
analysis of the pickup process in detail at the kinetic level.
Global models usually assume immediate assimilation of pickup ions
into the solar wind (one-fluid model) or perfect co-moving of
these populations $V_p = V_{pui}$ and no exchange of energy
$Q_{pui,p} =0$ (two- or three-fluid models). Energy exchange term
of pickup with ACRs,  $Q_{pui,acr} = -Q_{acr,pui}$, is specified
in next subsection.

\subsection{Cosmic rays}

The cosmic rays are coupled to background flow via scattering with
plasma waves. The net effect is that the cosmic rays tend to be
convected along with the background plasma as they diffuse through
the magnetic irregularities carried by the background plasma. Both
galactic and anomalous cosmic rays can be treated as populations
with negligible mass density and sufficient energy density. At a
hydrodynamical level, the cosmic rays may modify the wind flow via
their pressure gradient $\nabla P_c$ with the net energy transfer
rate
 from fluid to the cosmic rays given by $ \vec{V} \cdot \nabla P_c$.
$ P_{c}(\vec{r},t) = \frac{4 \pi}{3}
\int_0^{\infty} f_c(\vec{r}, p, t) wp^3 dp $ is a cosmic ray pressure;
$f_c(\vec{r}, p, t)$ is the isotropic velocity distribution of cosmic rays.

The transport equation of these particles has the following form \cite{Fichtner_2001}:
\begin{equation}\label{cr-f}
\frac{\partial f_c}{\partial t} = \frac{1}{p^2}
\frac{\partial}{\partial p} \left( p^2 D \frac{\partial
f_c}{\partial p}\right) + \nabla ( \widehat{k} \nabla f_c) - \vec{V}
\cdot \nabla f_c + \frac{1}{3} (\nabla \cdot \vec{V})
\frac{\partial f_c}{\partial ln p} + S(\vec{r}, p, t)
\end{equation}
Here $p$ is the modulus of the momentum of the particle; $D$ is
the diffusion coefficient in momentum space, often assumed to be
zero;
 $\widehat{k}$ is the tensor of spatial diffusion;
$\vec{V} = \vec{U} + \vec{V}_{drift}$ is the convection velocity;
$\vec{U}$ is the plasma bulk velocity; $\vec{V}_{drift}$ is a
drift velocity in the heliospheric or interstellar magnetic field;
and $S(\vec{r}, p,t)$ is the source term.

At the hydrodynamic level, the transport equation of the cosmic
rays in the heliospheric interface is:
\begin{equation}\label{cr-p}
\frac{\partial P_c}{\partial t} = \nabla [ \widehat{k} \nabla P_c -
\gamma_c ( \vec{U} + U_{dr}) P_c ] + (\gamma_c -1) \vec{U} \cdot
\nabla P_c + Q_{acr,pui}(\vec{r},t)
\end{equation}
Here we assume that $D=0$; $U_{dr}$ is momentum-averaged drift
velocity; $\gamma$ is the polytropic index; and $Q_{acr,pui}$ is
the energy injection rate describing energy gains of the ACRs from
pickup ions. Chalov and Fahr (\cite{chalov_fahr96},
\cite{chalov_fahr97}) suggested that $Q_{acr,pui} = - \alpha
p_{pui} div \vec{U} $, where $\alpha$ is a constant injection
efficiency defined by the specific plasma properties
\cite{chalov_fahr97}. $\alpha$ is set to zero for GCRs since no
injection occurs into the GCR component.

\section{OVERVIEW OF HELIOSPHERIC INTERFACE MODELS}
Together with Maxwellian equations (\ref{Maxwell}) the Boltzman
equation (\ref{eqBoltz}) for interstellar H atoms; sets of
hydrodynamic equations (\ref{density})-(\ref{energy}) written for
solar protons, electrons and pickup ions; and equation
(\ref{cr-p}) written for anomalous and cosmic ray components form
a closed system of equations, when interaction terms
$\vec{R}_{sr}$, $Q_{sr}$ are specified. Possible specification of
the interaction terms is given in equations (\ref{R}), (\ref{Q}).
We have to note here, that such a complete model has yet to be
developed. However, in recent years, several groups have focused
their efforts on theory and modeling in order to understand some
effects separately from others. In particular, the influence of
the interstellar magnetic field on the interface structure was
studied in \cite{Fujimoto_Matsuda_1991}- \cite{Myasnikov_1997} for
the two-dimensional case and in \cite{Ratkiewicz_1998} and
\cite{Pogorelov_Matsuda_1998} for the three-dimensional case. Both
interstellar and interplanetary magnetic fields were considered in
\cite{Linde} and \cite{Tanaka_Washimi_1999}. A comparison of these
MHD models was given recently in \cite{Ratkiewicz_2000}.
Latitudinal variations of the solar wind have been considered in
\cite{Pauls_Zank_1997}. The influence of the solar cycle
variations on the heliospheric interface was studied in the 2D
case in \cite{Steinolfson} - \cite{wang_99} and in
\cite{Tanaka_Washimi_1999} for the 3D case. In spite of many
interesting findings in the papers cited above, these theoretical
studies did not take into account the interstellar H atoms, or
took them into account but under greatly simplified assumptions,
as it was done in \cite{Linde}, where velocity and temperature of
interstellar H atoms were assumed as constants in the entire
interface.

Since most of the observational information on the heliospheric
interface is connected with interstellar neutrals and their
derivatives as pickup ions and ACRs, we will focus on the models,
which include interstellar neutrals in a more appropriate way.
These models can be separated into two types. Models of the first
type (Table 5) use a simplified fluid (or multi-fluid) approach
for interstellar H atoms. A kinetic approach was used in the
models of the second type. Development of the fluid (or
multi-fluid) models of H atoms was connected with the fact that
fluid (or multi-fluid) approach is simpler for numerical
realization. At the same time such an approach can lead to
nonphysical results. Results of one of the most sophisticated
multi-fluid models \cite{Zank_1996} were compared with the kinetic
Baranov-Malama model in \cite{bim98}. The comparison shows
qualitative and quantitative disagreements in distributions of H
atoms. At the same time, it was concluded in \cite{Williams_1997}
that the two models agreed on the distances to the termination
shock, heliopause and bow shock in upwind, but not in positions of
the termination shock in downwind.

\begin{sidewaystable}
\caption{Models with multi-fluid approaches for interstellar H atoms.} \label{table:models}
\newcommand{\m}{\hphantom{$-$}}
\newcommand{\cc}[1]{\multicolumn{1}{c}{#1}}
\renewcommand{\arraystretch}{1.2} 
\begin{tabular*}{\textheight}{@{\extracolsep{\fill}}lllllllllllll}
\hline & Reference   &  GCR & ACR & IMF & HMF & Latitud. SW & Time  & Pickup and & H atoms  &\\
& & & & & & asymmetry & Depend. & SW protons & & \\
\hline
& MULTI-FLUID &  APPROACH & FOR & H ATOMS &  & & & & & \\
& Liewer et al., 1995 \cite{Liewer} & $-$  & $-$ & $-$ & $-$ & $-$
&$+$ & one-fluid & one-fluid & \\
& Zank et al., 1996 \cite{Zank_1996} & $-$  & $-$ & $-$ & $-$ & $-$ &$-$ & one-fluid & three-fluid  & \\[2pt]
& Pauls and Zank, 1997 \cite{Pauls_Zank_1997} & $-$  & $-$ & $-$ & $-$ & $+$ &$-$ & one-fluid & one-fluid & \\[2pt]
& McNutt et al., 1998, 1999 \cite{mcnutt98}, \cite{mcnutt99} & $-$
& $-$ & $+$ & $+$ & $+$ & * & one-fluid &  one-fluid & \\[2pt]
& Wang and Belcher, 1999 \cite{wang_99} & $-$ & $-$ & $-$ & $-$ & $-$ & $+$ & one-fluid &  one-fluid & \\[2pt]
& Fahr et al., 2000 \cite{fahr00} & $+$ & $+$ & $-$ & $-$ & $-$ & $-$ & two-fluid
&  one-fluid & \\[2pt]
\hline
&KINETIC  &APPROACH & FOR  & H ATOMS & & & & & \\
&Osterbart and Fahr, 1992 \cite{osterbart} & $-$ & $-$ &$-$ &$-$ &$-$ &$-$ & No pickup
ions & not self- & \\
 & & & & & & & & & consistent & \\
&Baranov and Malama, 1993 \cite{bm93} & $-$ & $-$ & $-$ & $-$ & $-$ &$+$ & one-
fluid & Monte Carlo & \\
 & & & & & & & & & with splitting & \\
& Muller et al., 2000  \cite{Muller_2000} & $-$  & $-$ & $-$ & $-$ & $-$ &$-$ & 0ne-fluid & particle  & \\[2pt]
 & & & & & & & & & mesh code & \\
&Myasnikov et al, 2000 \cite{maic00}  & $+$ & $-$ & $-$ & $-$ & $-$ &$-$ & one-
fluid & Monte Carlo & \\
 & & & & & & & & & with splitting & \\
&Aleksashov et al, 2000 \cite{Aleksashov_2000}  & $-$ & $-$ & $+$ & $-$ & $-$ &$-$ & one-fluid & Monte Carlo & \\
 & & & & & & & & & with splitting & \\
&Zaitsev and Izmodenov, 2001 \cite{Zaitsev_Izmodenov_2001}  & $-$ & $-$ & $-$ & $-$ & $-$ &$+$ & one-fluid & Monte Carlo & \\
 & & & & & & & & & with splitting & \\
\hline
\end{tabular*}\\[2pt]
\end{sidewaystable}

\subsection{One-fluid plasma models}

One of the common features in the models \cite{wang_99},
\cite{Zank_1996}, \cite{Liewer} - \cite{mcnutt99}, \cite{bm93},
\cite{Muller_2000}, \cite{mica00}-\cite{Aleksashov_2000} is that
proton, electron and pickup ion components were considered as one
fluid. The great advantage of this approach is that its equations
are considerably simpler than the three- and two-fluid approaches
(see next subsection). A key assumption of this approach is
immediate assimilation of pickup protons into the original solar
protons. In other words, it is assumed that immediately after
ionization one cannot distinguish between original solar protons
and pickup protons. Another important assumption is that electron
and proton components have equal temperatures, $T_e = T_p$. For
quasineutral plasma ($n_p+n_{pui} = n_e + o(n_e)$) this means that
the pressure of the electrons is equal to half of total pressure
($P= n_e k T_e + (n_p+n_{pui}) k T_p \approx 2 n_e k T_e = 2
P_e$). Let us denote total density $\rho = m_e n_e
+m_p(n_p+n_{pui})$, bulk velocity $\vec{V} = (\sum_s m_s n_s
\vec{V}_s) / \rho$ ($s= e, p, pui$).

Governing equations for the one-fluid approach can be obtained by
summarizing equations (\ref{density})-(\ref{energy}) for indexes
$s = e, p, pui$. Introducing solar wind protons and pickup ions as
co-moving and taking into account that $m_e << m_p$ yield
one-fluid equations in their general form:
\begin{equation}\label{total_density}
\frac{\partial \rho}{\partial t} + \nabla \cdot (\rho \vec{V}) =
q_1, q_1= m_p n_H (\nu_{ph}+\nu_{impact})
\end{equation}

\begin{equation}\label{total_momentum}
\frac{\partial (\rho \vec{V}) }{\partial t} + \nabla P + \nabla
\cdot  (\rho \vec{V} \otimes \vec{V}) - \rho_e \vec{E} -
\frac{1}{c} [\vec{j} \times \vec{B}] = \vec{q}_2 - \nabla \cdot
n_e m_e \vec{u}_e \vec{u}_e
\end{equation}
Here $\vec{j} = \sum_\alpha n_\alpha e_\alpha \vec{V}_\alpha$,
$\rho_e = \sum_\alpha n_\alpha e_\alpha$; $\vec{q}_2 =
\sum_s m_s \vec{q}_{2,s}$; $\vec{u}_s = \vec{V}_s - \vec{V}$.
\begin{eqnarray}\label{total_energy}
\frac{\partial}{\partial t} \left( \frac{3}{2} P \right)  + \nabla
\cdot \left( \frac{5}{2} P \vec{V} \right) - \vec{V} \cdot \nabla
P = - \nabla \left( \frac{5}{2} P_e \vec{u}_e \right) + q_3-
\vec{q}_{2} \cdot \vec{V} + \\
(\vec{j}- \rho_e \vec{V}) \cdot \left(
\vec{E} + \frac{1}{c} \vec{V}_e \times \vec{B} +
\frac{m_e}{e n_e} (\vec{q}_{2,p} + \vec{q}_{2,pui} )
\right) \nonumber
\end{eqnarray}
where $q_3 = \sum_s m_s q_{3,s}$. The relative electron velocity
is connected with vector $\vec{j}$ as follows: $ \vec{j} = \rho_e
\vec{V} - e n_e \vec{u}_e$. Note that to derive
(\ref{total_energy}) we use a generalized form of Ohm's law and
neglect the terms proportional $m_e/m_p$ in it:
\[
n_e e \left( \vec{E} + \frac{1}{c} \vec{V}_e \times \vec{B}
\right) = - \nabla P_e + \vec{R}_{ep} + m_e (\vec{q}_{2,p}+
\vec{q}_{2,pui} - \vec{q}_{2,e})
\]
If we assume momentum transfer term $\vec{R_{ep}}$ as in (\ref{R})
and neglect the term $\rho_e \vec{V}$ in quasineutral plasma,
Ohm's law may be rewritten:
\begin{eqnarray}\label{Ohm}
\vec{E} = - \frac{1}{c} [\vec{V} \times \vec{B}]
+ \frac{m_e}{\tau_e n_e e^2} \vec{j}  + \frac{1}{e n_e} \left( \frac{1}{c} \vec{j}
\times \vec{B}  - \nabla P_e \right) + \frac{m_e}{e n_e} (\vec{q}_{2,p}+
\vec{q}_{2,pui} - \vec{q}_{2,e})
\end{eqnarray}

To derive the classical system of hydrodynamic equations applied
for heliospheric interface in one-fluid models, one needs to
ignore terms containing magnetic and electric fields in equations
(\ref{total_momentum}) and (\ref{total_energy}). If we also
disregard the second-order term $ \nabla n_e m_e \vec{u}_e
\vec{u}_e$ in (\ref{total_momentum}) and $\nabla (5p_e
\vec{u}_e/2)$ in (\ref{total_energy}), equations
(\ref{total_density})-(\ref{total_energy}) together with kinetic
equation (\ref{eqBoltz}) for H atoms form a closed system of
equations.

To get from (\ref{total_density})-(\ref{Ohm}) a system of ideally
conducting MHD equations with a magnetic field frozen in plasma,
one obviously needs to make other assumptions in addition to that
of high conductivity, when $R_m >>1$. $R_m = 4 \pi \sigma V L/c^2$
is magnetic Reynolds number, with the electrical conductivity
$\sigma = n_e e^2 \tau_e/m_e$, and $V$ and $L$ characteristic
velocity and length, respectively. Vanishing electron pressure
gradients, the Hall term, $\vec{j} \times \vec{B}$ and last term
of (\ref{Ohm}) connected with the charge-exchange effect,
corresponding terms in the generalized Ohm's law (\ref{Ohm}) also
must be ignored. Under these conditions, the Ohm's law has its
classical form $\vec{E} = - \frac{1}{c} [\vec{V} \times \vec{B}]$,
and last term of heat flux equation (\ref{total_energy}) is equal
to $j^2/\sigma$ and can be neglected.
 Ideal MHD equations with source terms $q_1,
\vec{q}_2, q_3$, on the right-hand sides were considered in
\cite{Linde}, \cite{mcnutt98}, \cite{mcnutt99},
\cite{Aleksashov_2000}.

\subsection{Three- and two-fluid plasma models}

For solar wind, the one-fluid model assumes essentially that
wave-particle interactions are sufficient for pickup ions to
assimilate quickly into the solar wind, becoming indistinguishable
from solar wind protons. However, as discussed above, Voyager
observations have shown that this is probably not the case. Pickup
ions are unlikely to be assimilated completely. Instead, two
co-moving thermal populations can be expected. A model that
distinguishes the pickup ions from the solar wind ions was
suggested by Isenberg in \cite{isenberg86}. Electrons were
considered as a third fluid. The key assumption in the model is
that pickup ions and solar wind protons are co-moving ($V_p =
V_{pui}$). It was also assumed that there is no exchange of
thermal energy between solar wind protons and pickup ions.
Isenberg's approach consists of two continuity equations
(\ref{density}) for solar protons and pickup ions; one momentum
equation (\ref{total_momentum}) and three energy equations
(\ref{energy}) for solar wind protons, electrons and pickup ions.
In (\ref{total_momentum}) Isenberg neglect term $\rho_e \vec{E}$,
the last term and assumes that $\vec{j} = c \nabla \times \vec{B}
/ (4 \pi)$. In energy equations he disregards the energy exchange
terms $Q_{sr}$. Note that Isenberg used the simplified form of
source terms suggested in \cite{holzer72} and applied these
equations to the spherically symmetric solar wind upstream the
termination shock.

Another two-fluid approach to model solar wind protons and pickup
protons  was developed recently in \cite{fahr00}. This model also
assumes that convection speed of pickup ions is identical to that
of solar wind protons. The pressure of pickup ions is calculated
by assuming a rectangular shape of the pickup ion isotropic
distribution function. In this case, the pressure can be expressed
through pickup ion density $\rho_{i}$ and solar wind bulk velocity
$V_{sw}$ as
\begin{equation}
P_{pui} = \rho_{pui} V_{sw}^2 /5.
\end{equation}
Therefore, governing plasma equations are i) one-fluid equations
for the mixture of solar protons, pickup ions and electrons; ii)
continuity equation for pickup ions; iii) two transport equations
for  ACRs and GCRs. The influence of cosmic ray components was
taken as terms $- \nabla (P_{ACR}+P_{GCR})$ and $- \vec{V} \cdot
\nabla (P_{ACR}+P_{GCR})- \alpha P_{pui} div(\vec{V})$ in the
right-hand side of the momentum and energy equations,
respectively.

\section{ BARANOV-MALAMA MODEL OF THE HELIOSPHERIC INTERFACE}

The first self-consistent model of the two-component (plasma and H atoms)
 LIC interaction with the solar wind was developed by Baranov and Malama \cite{bm93}.
The interstellar wind is assumed to have uniform parallel flow in the model.
The solar wind is assumed to be spherically symmetric at the Earth's orbit.
Under these assumptions, the heliospheric interface has axisymmetric structure.

Plasma and neutral components interact mainly by charge exchange.
However, photoionization, solar gravity and solar radiation
pressure, which are especially important in the vicinity of the
Sun, are also taken into account.

Kinetic and hydrodynamic approaches were used for the neutral and
plasma components, respectively. The kinetic equation
(\ref{eqBoltz} for neutrals is solved together with the Euler
equations for one-fluid plasma (\ref{total_density}) -
(\ref{total_energy}). The influence of the interstellar neutrals
is taken into account in the right-hand side of the Euler
equations that contain source terms $q_1$, $\vec{q}_2$, $q_3$,
which are integrals of the H atom distribution function
$f_H(\vec{V}_H)$ and can be calculated directly by the Monte Carlo
method \cite{malama91}. The set of kinetic and Euler equations is
solved by iterative procedure, as suggested in
\cite{Baranov_1991}. Supersonic boundary conditions were used for
the unperturbed interstellar plasma and for the solar wind plasma
at the Earth's orbit. The velocity distribution of interstellar
atoms is assumed to be Maxwellian in the unperturbed LIC. The
model results are discussed below in this section.

\begin{figure} \label{fig2}
\psfig{figure=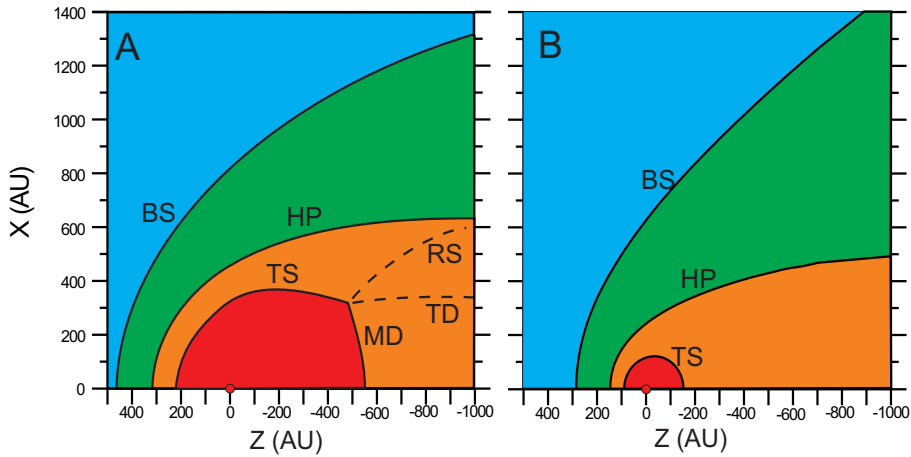,height=6cm,width=12cm,angle=0}
\caption{Effect of the interstellar neutrals on the size and
structure of the interface structure. (a) The heliospheric
interface pattern in the case of fully ionized local interstellar
cloud (LIC), (b) the case of partly ionized LIC. BS is the bow
shock. HP is the heliopause. TS is the termination shock. MD is
the Mach disk. TD is the tangential discontinuity and RS is the
reflected shock. }
\end{figure}

\begin{figure} \label{fig3}
\psfig{figure=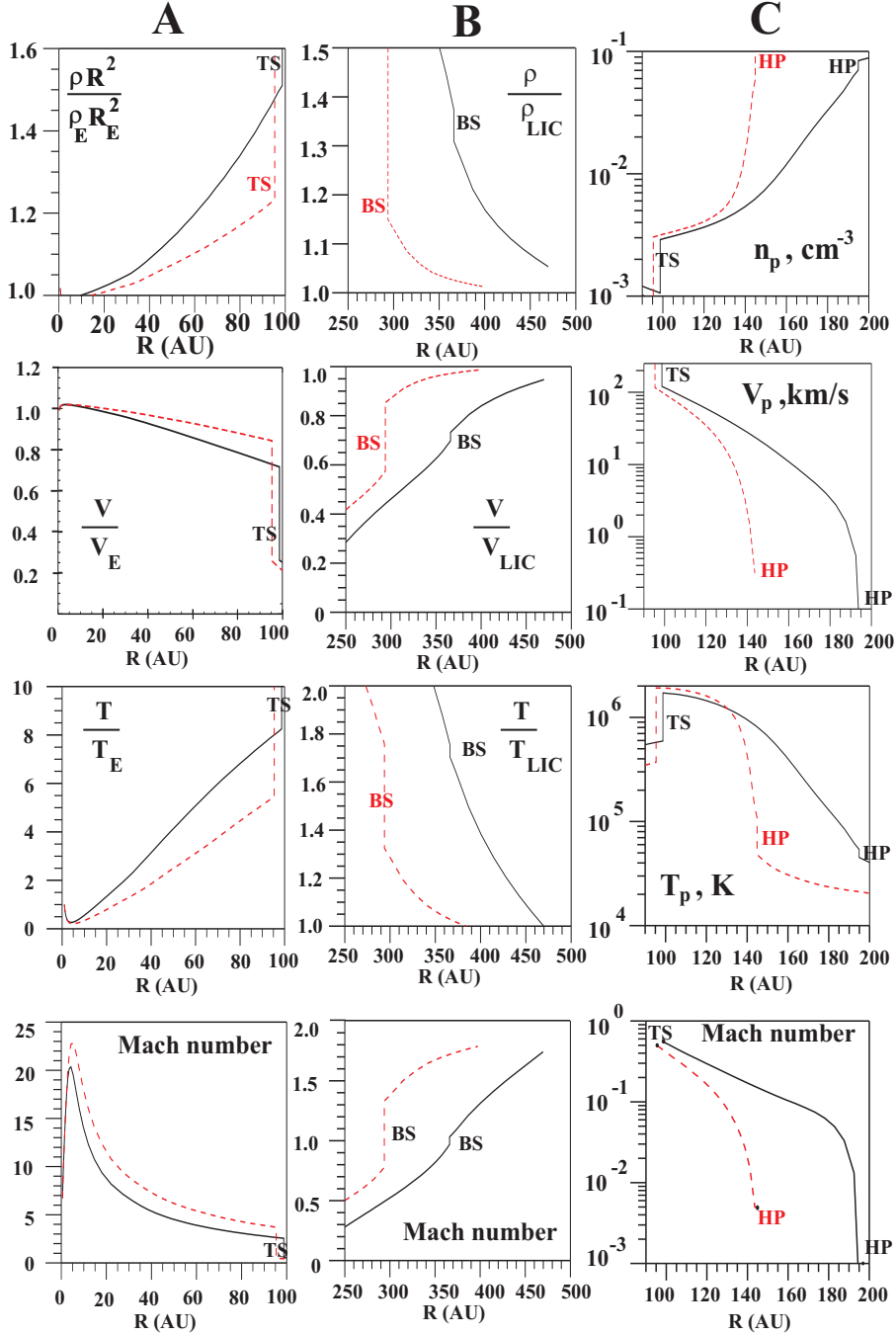,height=18cm,width=12cm,angle=0}
\caption{Plasma density, velocity, temperature and Mach number
upstream of the termination shock (A), upstream of the bow shock
(B), and in the heliosheath (C). The distributions are shown for
the upwind direction. Solid curves correspond to $n_{H,LIC}$=0.2
cm$^{-3}$, $n_{p,LIC}$=0.04 cm$^{-3}$. Dashed curves correspond to
$n_{H,LIC}$=0.14 cm$^{-3}$, $n_{p,LIC}$=0.10 cm$^{-3}$.
$V_{LIC}$=25.6 km/s, $T_{LIC}$=7000 K}
\end{figure}
\begin{figure} \label{fig4_5}
\begin{minipage}[t]{75mm}
\psfig{figure=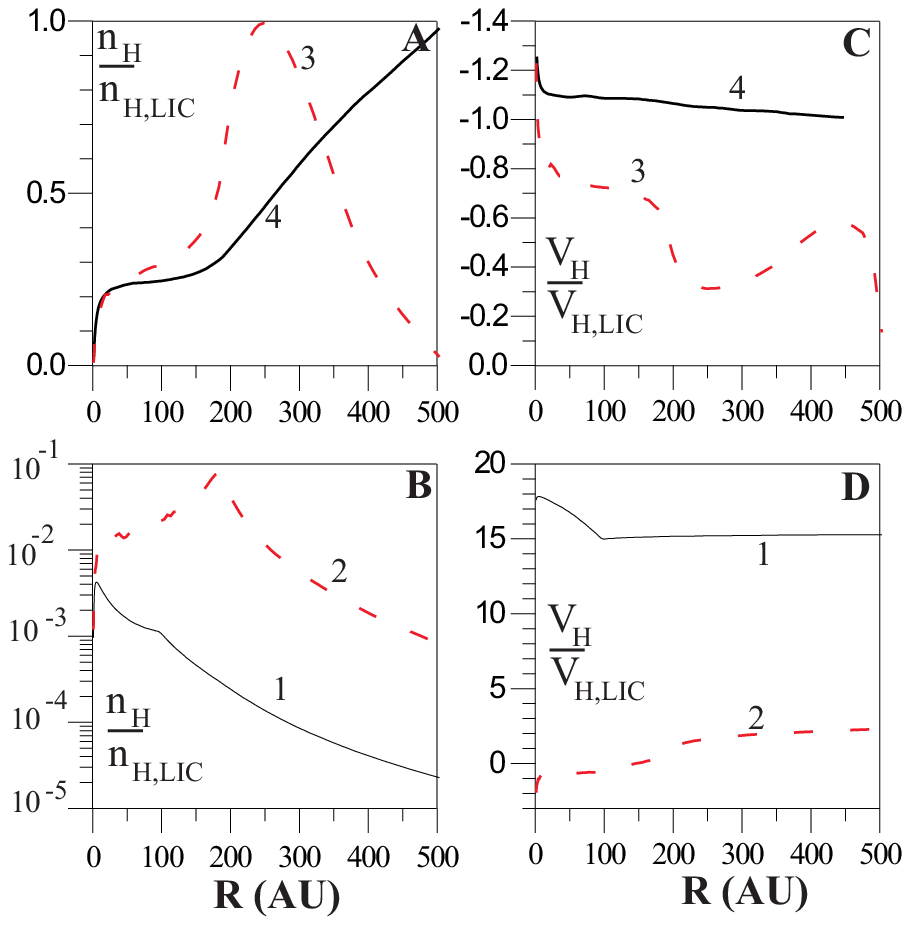,height=8cm,width=8cm,angle=0}
\caption{Number densities and velocities of 4 atom populations as
functions of heliocentric distance in the upwind direction. 1
designates atoms created in the supersonic solar wind, 2 atoms
created in the heliosheath, 3  atoms created in the disturbed
interstellar plasma, and 4 original (or primary) interstellar
atoms. Number densities are normalized to $n_{H,LIC}$, velocities
are normalized to $V_{LIC}$. It is assumed that $n_{H,LIC}=0.2$
cm$^{-3}$, $n_{p,LIC}=0.04$ cm$^{-3}$. }
\end{minipage}
\hspace{\fill}
\begin{minipage}[t]{75mm}
\psfig{figure=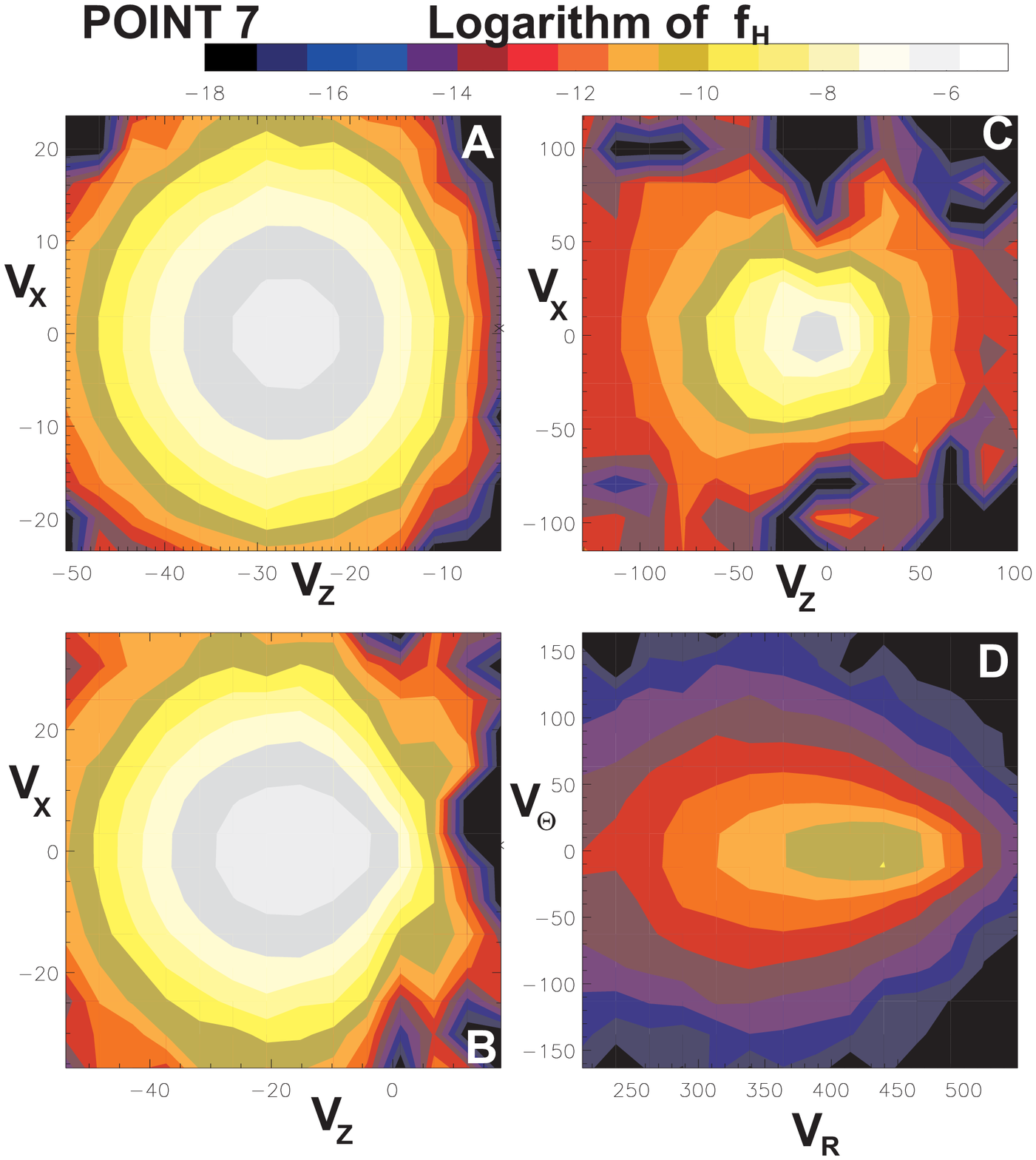,height=8cm,width=8cm,angle=0}
\caption{ Velocity distributions of four atom populations at the
termination shock in the upwind direction. (A) primary
interstellar atoms, (B) secondary interstellar atoms, (C) atoms
created in the heliosheath, (D) atoms created in the supersonic
solar wind. $V_z$ is the projection of velocity on the axis
parallel to the LIC velocity vector. Negative values of $V_z$
indicate approach to the Sun. $V_x$ is the radial component of the
projection of velocity vector on the perpendicular plane. $V_z$,
$V_x$ are in cm/sec. It is assumed that $n_{H,LIC}=0.2$ cm$^{-3}$,
$n_{p,LIC}=0.04$ cm$^{-3}$. }
\end{minipage}
\end{figure}
\subsection{Plasma}

Interstellar atoms strongly influence the heliospheric interface
structure. In the presence of interstellar neutrals, the
heliospheric interface is much closer to the Sun than in a pure
gas dynamical case (Figure 2). The termination shock becomes more
spherical. The Mach disk and the complicated shock structure in
the tail disappear.

The supersonic plasma flows upstream of the bow and termination
shocks are disturbed. The supersonic solar wind is disturbed by
charge exchange with the interstellar neutrals. The new ions
created by charge exchange are picked up by the solar wind
magnetic field. The Baranov-Malama model assumes immediate
assimilation of pickup ions into the solar wind plasma. The solar
wind protons and pickup ions are treated as one-fluid, called the
solar wind. The number density, velocity, temperature, and Mach
number of the solar wind are shown in Figure 3A. The effect of
charge exchange on the solar wind is significant. By the time the
solar wind flow reaches the termination shock, it is  decelerated
(15-30 \%), strongly heated (5-8 times) and mass loaded (20-50 \%)
by the pickup ion component.

The interstellar plasma flow is disturbed upstream of the bow
shock by charge exchange with the secondary atoms originating in
the solar wind and compressed interstellar plasma. Charge exchange
results in the heating (40-70 \%) and deceleration (15-30 \%) of
the interstellar plasma before it reaches the bow shock. The Mach
number decreases and for a certain set of interstellar parameters
($n_{H,LIC} >> n_{p,LIC}$) the bow shock may disappear. Solid
curves on Figure 3B correspond to the small ionization degree of
LIC ($ n_p / (n_p+n_H) = 1/6$). The bow shock almost disappears.

The interstellar neutrals also modify the plasma structure in the
heliosheath. In a pure gas dynamic case (without neutrals) the
density and temperature of the  postshock plasma are nearly
constant. However, the charge exchange process leads to a large
increase of the plasma number density and a decrease of its
temperature (Figure 3C). The electron impact ionization process
may influence the heliosheath plasma flow by increasing the
gradient of the plasma density from the termination shock to the
heliopause \cite{bm96}. The effects of interstellar atom influence
on the heliosheath plasma flow may be important, in particular,
for the interpretations of kHz radio emission detected by Voyager
(\cite{Gurnett_Kurth}, \cite{Treumann_1998}) and  possible future
heliospheric imaging in energetic neutral atom (ENA) fluxes
\cite{Gruntman_2001}.

\subsection{Atoms}

Charge exchange significantly disturbs the interstellar atom flow.
Atoms newly created by charge exchange have velocities of their
ion partners in charge exchange collisions. Therefore, the
velocity distribution of these new atoms depends on the local
plasma properties. It is convenient to distinguish four different
populations of atoms depending on where in the heliospheric
interface they originated. Population 1 is the atoms created in
the supersonic solar wind. Population 2 is the atoms originating
in the heliosheath. Population 3 is the atoms created in the
disturbed interstellar wind. We will call original (or primary)
interstellar atoms population 4. The number densities and mean
velocities of these populations are shown in Figure 4 as the
function of the heliocentric distance. The velocity distribution
function of interstellar atoms $f_H(\vec{w}_H, \vec{r})$ can be
represented as a sum of the distribution functions of these
populations: $ f_H = f_{H,1}+ f_{H,2} + f_{H,3} + f_{H,4} $. The
Monte Carlo method allows us to calculate these four distribution
functions. The velocity distributions of the interstellar atoms in
the 12 selected points in the heliospheric interface were
presented in \cite{Izmodenov_2001}. For example, the velocity
distributions at the termination shock in the upwind direction are
shown in Figure 5. Note that velocity distributions of H atoms in
the heliosphere were also presented in \cite{Muller_2000}.
However, different populations of H atoms cannot be considered
separately in mesh particle simulations of H atoms
\cite{Lipatov_1998}.

{\bf Original (or primary) interstellar atoms} are significantly
filtered (i.e. their number density is reduced) before reaching
the termination shock (Figure 4A). Since slow atoms have a smaller
mean free path as compared with fast atoms, they undergo more
charge exchange. This kinetic effect, called ``selection'',
results in a deviation of the interstellar distribution function
from Maxwellian ( Figure 5A). The selection also results in
$\sim$10 \% increase of the primary atom mean velocity to the
termination shock (Figure 4C).

{\bf The secondary interstellar atoms} are created in the
disturbed interstellar medium by charge exchange of primary
interstellar neutrals and protons decelerated by the bow shock.
The secondary interstellar atoms collectively make up the ``H
wall'', a density increase at the heliopause. The ``H wall'' has
been predicted in \cite{Baranov_1991} and detected toward $\alpha$
Cen \cite{Linsky_Wood96}. At the termination shock, the number
density of the secondary neutrals is comparable to the number
density of the primary interstellar atoms (Figure 4A, dashed
curve). The relative abundances of the secondary and primary atoms
entering the heliosphere vary with degree of interstellar
ionization. It has been shown in \cite{Izmodenov_1999} that the
relative abundance of the secondary interstellar atoms inside the
termination shock increases with increasing interstellar proton
number density. The bulk velocity of the population 3 is about -18
-19 km/s. The sign ``-'' means that the population approaches the
Sun. One can see that the velocity distribution of this population
is not Maxwellian (Figure 5B). The reason for the abrupt behavior
of the velocity distribution for $V_z >0$ is that the particles
with significant positive $V_z$ velocities can reach the
termination shock only from the downwind direction. The velocity
distributions of different populations of H atoms were calculated
in \cite{Izmodenov_2001} for different directions from upwind. The
fine structures of the velocity distribution of the primary and
secondary interstellar populations vary with direction. These
variations of the velocity distributions reflect the geometrical
pattern of the heliospheric interface. The velocity distributions
of the interstellar atoms can be a good diagnostics of the global
structure of the heliospheric interface.

The third population of the heliospheric neutrals is {\bf the
neutrals created in the heliosheath} from hot and compressed solar
wind protons. The number density of this population is an order of
magnitude smaller than the number densities of the primary and
secondary interstellar atoms. This population has a minor
importance for interpretations of Ly $\alpha$ and pickup ion
measurements inside the heliosphere. However, some of these atoms
may probably be detected by Ly $\alpha$ hydrogen cell experiments
due to their large Doppler shifts. Due to their high energies, the
particles influence the plasma distributions in the LIC. Inside
the termination shock the atoms propagate freely. Thus, these
atoms can be the source of information on the plasma properties in
the place of their birth, i.e. the heliosheath
{\cite{Gruntman_2001}.

The last population of heliospheric atoms is {\bf the atoms
created in the supersonic solar wind}. The number density of this
atom population has a maximum at $\sim$5 AU. At this distance, the
number density of population 1 is about two orders of magnitude
smaller than the number density of the interstellar atoms. Outside
the termination shock the density decreases faster than $1/r^2$
where $r$ is the heliocentric distance (curve 1, Figure 4B). The
mean velocity of population 1 is about 450 km/sec, which
corresponds to the bulk velocity of the supersonic solar wind. The
velocity distribution of this population is not Maxwellian either
(Figure 5D). The extended ``tail'' in the distribution function is
caused by the solar wind plasma deceleration upstream of the
termination shock. The ``supersonic'' atom population results in
the plasma heating and deceleration upstream of the bow shock.
This leads to the decrease of the Mach number ahead of the bow
shock.

\subsection{ Recent developments in the Baranov-Malama model}

The Baranov-Malama model, the basic results of which were
discussed above, takes into account essentially two interstellar
components: H atoms and charged particles. To apply this model to
space experiments, one needs to evaluate how other possible
components of the interstellar medium influence the results of
this two-component model. Recently, several effects were taken
into account in the frame of this axisymmetric model.

The influence of the galactic cosmic rays on the heliospheric
interface structure was studied recently in \cite{mica00},
\cite{maic00}. The study was done in the frame of two-component
(plasma and GCRs) and three-component (plasma, H atoms and GCRs)
models. For the two-component case it was found that cosmic rays
could considerably modify the shape and structure of the solar
wind termination shock and the bow shock and change the positions
of the heliopause and the bow shock. At the same time, for the
three-component model it was shown \cite{maic00} that the GCR
influence on the plasma flows is negligible as compared with the
influence of H atoms. The exception is the bow shock, a structure
that can be strongly modified by the cosmic rays. It was also
found (\cite{fahr00}; Alexashov, private communication) that an
anomalous component does not have a significant effect on the
position of the termination shock. However, ACRs may significantly
reduce compression at the termination shock \cite{fahr00}.

Effects of the interstellar magnetic field on the plasma flow and
on distribution of H atoms in the interface were studied in
\cite{Aleksashov_2000} in the case of magnetic field parallel to
the relative Sun/LIC velocity vector. In this case, the model
remains axisymmetric. It was shown that effects of the the
interstellar magnetic field on the positions of the termination
and bow shocks and the heliopause are significantly smaller as
compared to model with no atoms \cite{Baranov_Zaitsev_1995}. The
calculations were performed with various Alfven Mach numbers in
the undisturbed LIC. It was found that the bow shock straightens
out with decreasing Alfven Mach number (increasing magnetic field
strength in LIC). It approaches the Sun near the symmetry axis,
but recedes from it on the flanks. By contrast, the nose of the
heliopause recedes from the Sun due to tension of magnetic field
lines, while the heliopause in its wings approaches the Sun under
magnetic pressure. As a result, the region of compressed
interstellar medium around the heliopause (or "pileup region")
decreases by almost 30 $\%$, as the magnetic field increases from
zero to 3.5 $\times 10^{-6}$ Gauss. It was also shown in
\cite{Aleksashov_2000} that H atom filtration and heliospheric
distributions of primary and secondary interstellar atoms are
virtually unchanged over the entire assumed range of the
interstellar magnetic field (0 - $3.5 \cdot 10^{-6}$ Gauss). The
magnetic field has the strongest effect on density distribution of
population 2 of H atoms, which increases by a factor of almost 1.5
as the interstellar magnetic field increases from zero to $3.5
\cdot 10^{-6}$ Gauss.

Very recently a new non-stationary model of the solar wind
interaction with two-component (H atoms and plasma) LIC was
proposed in \cite{Zaitsev_Izmodenov_2001}. In this model the
primary and secondary interstellar atoms (populations 3 and 4)
were treated as quasi-stationary kinetic gases. Population 1 of
atoms originating in the supersonic solar wind was considered as
zero-pressure fluid. The calculations show that the qualitative
features of the non-stationary SW/LIC interaction established in
\cite{Baranov_Zaitsev_1998} remain, but the effect of the solar
activity cycle is quantitatively stronger because the interface is
closer to the Sun than in the model with no atoms. The motion of
the termination shock during the solar cycle on the axis of
symmetry is about 30 AU. Due to the solar cycle variations of the
neutralized solar wind (i.e. atoms of population 1) the region
between the heliopause and the bow shock widens and the mean
plasma density in the region becomes smaller than for the
stationary problem.

\section{INTERPRETATIONS OF SPACECRAFT EXPERIMENTS ON THE BASIS OF THE BARANOV-MALAMA MODEL}

The Sun/LIC relative velocity and the LIC temperature are now well
constrained (\cite{Witte_1996},
\cite{Lallement_Bertin}-\cite{Lallement_1995}). Using the SWICS
pickup He results and an interstellar HI/HeI ratio of 13 $\pm$ 1
(the average value of the ratio toward the nearby white dwarfs),
Gloeckler et al.\cite{Gloeckler_1997} concluded that $n_{LIC}(HI)
= 0.2 \pm 0.03$ cm $^{-3}$. This estimation of $n_{LIC}(HI)$ is
independent of the heliospheric interface model but
model-dependent for determination of the number density of H atoms
from pickup fluxes. Estimates of interstellar electron number
density require a theoretical model of the heliospheric interface.
The Baranov-Malama model was used in \cite{Izmodenov_1999} to
study the sensitivity of the various types of indirect diagnostics
of local interstellar plasma density. The diagnostics are the
degree of filtration, the temperature and the velocity of the
interstellar H atoms in the outer heliosphere (at the termination
shock), the distances to the termination shock, the heliopause,
and the bow shock, and the plasma frequencies in the LIC, at the
bow shock and in the maximum compression region around the
heliopause, which constitutes the ``barrier'' for radio waves
formed in the interstellar medium. We also searched
\cite{Izmodenov_1999} for a number density of interstellar protons
compatible with SWICS/Ulysses pickup ion observations,
backscattered solar Ly $\alpha$ observed by SOHO, Voyager and HST,
and kHz radiations observed by Voyager. Table 1 presents the
ranges of $n_{p, LIC}$ obtained on the basis of the Baranov-Malama
model and comparable to these observations.

\begin{table}[hbt]
  \centering
\caption{Intervals of Possible Interstellar Proton Number Densities}
\begin{tabular*}{\textwidth}{@{}l@{\extracolsep{\fill}}ll}
\hline
Type of Heliospheric Interface Diagnostics  & Range of Interstellar Proton Number Density  \\
\hline
SWICS/Ulysses pick-up ion \cite{Gloeckler_1997}               & \\
\hspace{4pt} 0.09 cm$^{-3} < n_{\rm H,TS} < 0.14$ cm$^{-3}$  &  0.02 cm$^{-3}< n_{\rm p, LIC} < 0.1$ cm$^{-3} $  \\
Ly- $\alpha$,  intensity \cite{Quemerais_1994}                  & \\
\hspace{4pt}  0.11 cm$^{-3} < n_{H, TS} < 0.17$ cm$^{-3}$ & $ n_{\rm p, LIC} < 0.04$ cm$^{-3} $ or \\
Ly-$\alpha$, Doppler shift \cite{Bertaux_1985} - \cite{Clarke}              &                 \\
\hspace{4pt} 18 km s$^{-1} < V_{H, TS} < 21$ km s$^{-1} $    &  0.07 cm$^{-3} <n_{\rm p, LIC} < 0.2$ cm$^{-3} $    \\
Voyager kHz emission (events) \cite{Gurnett_Kurth}      &   \\
\hspace{4pt}  110 AU $< R_{\rm AU} <$ 160 AU  &  0.08 cm$^{-3} < n_{\rm p, LIC} < 0.22$ cm$^{-3} $  \\
Voyager  kHz emission (cutoff) \cite{Gurnett_1993}, \cite{Grzedzielski_Lallement_1996}      &   \\
\hspace{5pt}    1.8 kHz                        & $  n_{\rm p, LIC} = 0.04$ cm$^{-3}$ \\
\hline
\end{tabular*}
\end{table}

From analysis of the ranges, it was concluded in
\cite{Izmodenov_1999} that it is difficult in the frame of the
model to reconcile the results obtained from all types of data as
they stand now. There is a need for some modifications of the
interpretations or of the confidence intervals. Two mutually
exclusive solutions have been suggested: (1) It is possible to
reconcile the pickup ions and Ly $\alpha$ measurements with the
radio emission time delays if a small additional interstellar
(magnetic or low-energy cosmic ray) pressure is added to the main
plasma pressure. In this case, $ n_{\rm p, LIC}= 0.07$ cm$^{-3} $
and $ n_{\rm H, LIC}= 0.23$ cm$^{-3} $ is the favored pair of
interstellar densities. However, in this case, the low frequency
cutoff at 1.8 kHz does not correspond to the interstellar plasma
density, and one has to search for another explanation. (2) The
low-frequency cutoff at 1.8 kHz constrains the interstellar plasma
density, i.e., $ n_{\rm p,LIC}=0.04$ cm$^{-3} $. In this case, the
bulk velocity deduced from the Ly $\alpha$ spectral measurement is
underestimated by about 30-50\% (the deceleration is about 3 km
s$^{-1}$ instead of 5-6 km s$^{-1}$). Model limitations (e.g. a
stationary hot model to derive the bulk velocity) or the influence
of a strong solar Ly $\alpha$ radiation pressure may play a role.
In this case, a significant additional interstellar (magnetic or
cosmic ray) pressure as compared with case (1) would be needed.

This need for an additional pressure is in agreement with the
conclusions made in \cite{Gayley_1997}, which were derived from
the analysis of the H wall absorption toward alpha Centauri
\cite{Linsky_Wood96}. In their model the authors modified the
equation of the state of the gas to simulate the effect of the
interstellar magnetic field (IMF) and concluded that H wall
absorption favors the ``subsonic case''. However, the best model
of these authors corresponds to a neutral H density of 0.025
cm$^{-3}$ in the inner heliosphere, at least 4 times smaller than
the density derived from the pickup ions. Also, the precision
required to model the differences between the theoretical
absorptions, namely small differences of the order of a few
kilometers per second at the bottom of the lines, is of the order
of the differences between the kinetic and multi-fluid model
results for the same parameters in the supersonic case (see
Appendix B in \cite{Williams_1997} ; \cite{bim98}). Thus, an
additional study of the absorption toward nearby stars for more
realistic densities and models is desired.

\section{PROBLEMS FOR FUTURE WORK}

The Local Interstellar Medium interacts with the solar wind and
influences the outer heliosphere in a complicated way. Several
particle populations and magnetic fields are involved in this
interaction. From the interstellar side, the interacting
populations are the plasma (electron and proton) component, H atom
component, interstellar magnetic field, and galactic cosmic rays.
Heliospheric plasma consists of original solar wind protons,
electrons, pickup protons, and the anomalous component of cosmic
rays. A large effort has been done to study the theoretical
physics of the interaction region. However, a complete,
self-consistent model of the heliospheric interface has not yet
been constructed, because of the difficulty connecting both the
multi-fluid nature of the heliosphere and the requirements of the
different theoretical approaches for different components of the
interaction. Many aspects were studied and reported here in
previous sections. However, some aspects require additional
theoretical explorations. Most theoretical models employ the
one-fluid approach for solar wind and interstellar plasmas. It has
been shown that, to derive one-fluid approach equations, several
assumptions are needed. A key assumption that looks reasonable  is
co-moving character of all components. Another assumption for a
one-fluid plasma model is the immediate assimilation of the pickup
ion component into the solar wind. As demonstrated by space
experiments, this is not the case and it would be more natural to
consider solar protons and pickup protons separately as co-moving
populations. The electron component should also be treated as a
distinct population. However, since the assumption of the
co-moving character of these three heliospheric plasma populations
looks reasonable, the one-fluid approach gives us a reasonably
accurate picture of the flow pattern (positions of the shocks and
heliopause) and plasma velocity distributions. Theoretical models
of pickup ion acceleration and diffusion can be employed to
determine the distribution of thermal energy between solar wind
and pickup proton components. A similar study should be done for
electrons.

Another important aspect of the solar/wind interaction is a study
of the tail region of the solar wind and interstellar medium
interaction. Although some studies were done
(\cite{Jaeger_Fahr_1998}, \cite{fahr_86}) it is still not clear at
which heliocentric distances the gas (plasma and H atoms)
parameters become indistinguishable from local interstellar
parameters, or in other words, how far signatures of the solar
system are noticeable in the interstellar medium. It is still not
clear which of the two competiting processes is the most important
in the tail region - charge exchange or plasma transport across
the heliopause due to different instabilities. Studies of Saturn's
and Earth's magnetic tails show that such tails can be very
extended \cite{Grzedzielski_81}, \cite{Grzedzielski_88}.

Finally, growing interest in heliospheric interface studies is
connected with expectations that Voyager 1 will cross the
termination shock soon. Many predictions of the time of the
termination shock crossing by Voyager appeared in the literature.
However, it seems that much more work should be done to explain
and reconcile all available indirect observations of the
heliospheric interface based on the unique model of the
heliospheric interface. This work should be done especially
because NASA plans to send a spacecraft to a heliocentric distance
of at least 200 AU with a flight-time of only 10 or 15 years.
Intensive theoretical study will help to optimize goals,
instrumentation, and, finally, the scientific profit of this
"interstellar" mission.

{\bf Acknowledgements.} This work was supported in part by CRDF
Award RP1-2248, INTAS Award  2001-0270, YSF 00-163, RFBR grants
01-02-17551, 02-02-06011, 01-01-00759, and the International Space
Science Institute in Bern. I thank V.~B. Baranov and S.~V. Chalov
for useful discussions.


\begin{thebibliography}{}
\bibitem{issibook_96} R. von Steiger, R. Lallement, and M. A. Lee (eds.),
The Heliosphere in the Local Interstellar Medium, Hardbound, 1996.
\bibitem{Linsky_Wood96} Linsky, J., Wood, B., Astrophys. J. 463 (1996), 254.
\bibitem{Wood2000} Wood, B. E., Muller, H.; Zank, G. P.,
Astrophys. J. 542 (2000), 493-503.
\bibitem{ilm99} Izmodenov, V., Lallement, R., Malama, Y., Astron. Astrophys. 342 (1999), L13-L16.
\bibitem{iwl2002} Izmodenov, V., Wood, B., Lallement, R., J. Geophys. Res., in press, 2002.
\bibitem{Gruntman_2001} Gruntman et al., J. Geophys. Res. 106, 15767-15782 (2001).
\bibitem{parker61} Parker, E. N., Astrophys. J. 134 (1961), 20-27.
\bibitem{baranov71} Baranov, V.B., Krasnobaev, K.V., Kulikovksy, A.G.,
Sov. Phys. Dokl. 15 (1971), 791.
\bibitem{Fichtner_2001} Fichtner, H., Space Sci. Rev. 95 (2001), 639-754.
\bibitem{Zank_1999} Zank, G.,  Space Sci. Rev. 89 (1999), 413-688.
\bibitem{Richardson_2001}Richardson, J.D.,
The Outer Heliosphere: The Next Frontiers, Edited by K. Scherer,
H. Fichtner, H. Fahr, and E. Marsch, COSPAR Colloquiua Series, 11.
Amsterdam: Pergamon Press (2001), 301-310.
\bibitem{Lallement_1996}Lallement, R., Space Sci. Rev. 78 (1996), 361-374.
\bibitem{Witte_1996} Witte, M., Banaszkiewicz, M.; Rosenbauer, H.,
Space Sci. Rev. 78 (1996), 289-296.
\bibitem{Gloeckler_1996} Gloeckler, Space Sci. Rev. 78 (1996), 335-346.
\bibitem{Moebius_1996}Moebius, E., Space Sci. Rev. 78 (1996), 375-386.
\bibitem{Baranov_Malama_1995} Baranov, V.~B., Malama, Y.~G., J. Geophys. Res. 100 (1995), 14,755-14,762.
\bibitem{Izmodenov_1999} Izmodenov V., Geiss, J., Lallement, R., et al.,
J. Geophys. Res. (1999), 4731-4742.
\bibitem{Webber_2001} Webber, W.R., Lockwood, J., McDonald, F., Heikkila, B.,
J. Geophys. Res. 106 (2001), 253-260.
\bibitem{Gloeckler_1997} Gloeckler, Nature 386 (1997), 374-377.
\bibitem{izmod_et_al_2000} Izmodenov, V., Malama, Y., Kalinin, A.,et al., Astrophys. Space Sci. 274 (2000), 71-76.
\bibitem{baranov2000} Baranov, V.~B., Astrophys. Space Sci. 274 (2000), 3-16.
\bibitem{braginski} Braginski, S.I., Voprosy teorii plasmy, v.1, Atomizdat, Moscow, 1963 (in russian).
\bibitem{isenberg97} Isenberg, P., J. Geophys. Res. 102 (1997), 4719-4724.
\bibitem{chalov_fahr98}Chalov, S.~V., Fahr, H., Astron. Astrophys., 335 (1998), 746-756.
\bibitem{Williams_1995} Williams, L., Zank, G., Matthaeus, W., J. Geophys. Res. 100 (1995), 17059-17068.
\bibitem{chalov_fahr96}Chalov, S.~V., Fahr, H., Astron. Astrophys. 311 (1996), 317-328.
\bibitem{chalov_fahr97}Chalov, S.~V., Fahr, H., Astron. Astrophys. 326 (1997), 860-869.
\bibitem{Fujimoto_Matsuda_1991} Fujimoto Y., Matsuda, T., Preprint No. KUGD91-2,
Kobe Univ., Japan, 1991.
\bibitem{Baranov_Zaitsev_1995} Baranov, V.B., Zaitsev, N.A., Astron. Astrophys. 304 (1995), 631.
\bibitem{Pogorelov_Semenov_1997} Pogorelov, N., Semenov, A., Astron. Astrophys. 321 (1997), 330.
\bibitem{Myasnikov_1997} Myasnikov, A., Preprint No. 585, Institute for Problems in Mechanics, Russian Academy of Sciences, 1997.
\bibitem{Ratkiewicz_1998} Ratkiewicz, R., Barnes, A, et al., Astron. Astrophys. 335 (1998), 363.
\bibitem{Pogorelov_Matsuda_1998} Pogorelov, N., Matsuda, T., J. Geophys. Res. 103 (1998), 237-245.
\bibitem{Linde} Linde, T., Gombosi, T., Roe, P., J. Geophys. Res. 103 (1998), 1889-1904.
\bibitem{Tanaka_Washimi_1999} Tanaka, T., Washimi, H.,  J. Geophys. Res. 104 (1999), 12605.
\bibitem{Ratkiewicz_2000} Ratkiewicz, R., Barnes, A., J. Spreiter, J. Geophys. Res. 105 (2000), 25,021-25,031.
\bibitem{Pauls_Zank_1997} Pauls, H. and G. Zank, J. Geophys. Res. 102 (1997), 19779-19788.
\bibitem{Steinolfson} Steinolfson, R.S., J. Geophys. Res. 99 (1994), 13,307-13,314.
\bibitem{Pogorelov_1995} Pogorelov, N.,  Astron. Astrophys. 297 (1995), 835.
\bibitem{Karmesin} Karmesin, S., Liewer, P, Brackbill, J., Geophys. Res. Let. 22 (1995), 1153-1163.
\bibitem{Baranov_Zaitsev_1998} Baranov, V., Zaitsev, N.,
Geophys. Res. Let. 25 (1998), 4051.
\bibitem{wang_99} Wang, C., Belcher, J, J. Geophys. Res. 104 (1999), 549-556.
\bibitem{Zank_1996} Zank, G., Pauls, H., Williams, L., Hall, D., J. Geophys. Res. 101 (1996), 21639-21656.
\bibitem{bim98} Baranov, V.~B., Izmodenov, V., Malama, Y., J. Geophys. Res. 103 (1998), 9575-9586.
\bibitem{Williams_1997}Williams, L., Hall, D.~T., Pauls, H.~L., Zank, G.~P., Astrophys. J. 476 (1997), 366.
\bibitem{Liewer} Liewer, P., Brackbill, J., Karmesin, S., International Solar Wind 8 Conference, p.33, 1995.
\bibitem{mcnutt98}  McNutt, R., Lyon, J., Goodrich, C., J. Geophys. Res. 103 (1998), 1905.
\bibitem{mcnutt99}  McNutt, R., Lyon, J., Goodrich, C., J. Geophys. Res. 104 (1999), 14803.
\bibitem{fahr00} Fahr, H., Kausch, T., Scherer, H., Astron. Astrophys. 357 (2000), 268-282.
\bibitem{osterbart}Osterbart, R., and  H. Fahr,  Astron. Astrophys. 264 (1992), 260-269.
\bibitem{bm93} Baranov, V., Malama, Y., J. Geophys. Res. 98 (1993), 15157.
\bibitem{Muller_2000} Muller et al., J. Geophys. Res., 27,419-27,438 (2000)
\bibitem{Lipatov_1998} Lipatov et al., J. Geophys Res., 1998.
\bibitem{bm96} Baranov, V.~B., and Y.~G. Malama, Space Sci. Rev. 78 (1996), 305-316.
\bibitem{Baranov_1991} Baranov, V.~B., Lebedev, M.,Malama Y., Astrophys. J. 375 (1991), 347-351.
\bibitem{malama91} Malama, Y.~G., Astrophys. Space Sci., 176 (1991), 21-46.
\bibitem{Gurnett_Kurth} Gurnett, D.,Kurth,W.,Space Sci. Rev. 78 (1996), 53-66.
\bibitem{Quemerais_1994} Quemerais, E., Bertaux, J.-L., Sandel, B., Lallement, R., Astron. Astrophys. 290 (1994), 941-955.
\bibitem{Bertaux_1985} Bertaux, J.-L., Lallement, R., Kurt, V., Mironova,E.~N.,
Astron. Astrophys. 150 (1985), 1-20.
\bibitem{Lallement_et_al_1996} Lallement, R., Linsky, J., Lequeux, J., Baranov, V., Space Sci. Rev. 78 (1996), 299-304.
\bibitem{Clarke} Clarke, J., Lallement, R., Quemerais, E., Bertaux, J.-L., Scherer, H., Astron. Astrophys. 499 (1998), 482.
\bibitem{Gayley_1997} Gayley, K., Zank G.~P., et al.,  Astron. Astrophys. 487 (1997), 259.
\bibitem{Gurnett_1993} Gurnett, D.~A., Kurth, W., Allendorf, S., Poynter, R., Science 262 (1993), 199-202.
\bibitem{Grzedzielski_Lallement_1996} Grzedzielski, S., Lallement, R., Space Sci. Rev. 78 (1996), 247-258.
\bibitem{Treumann_1998} Treumann, R., Macek, W., Izmodenov, V, Astron. Astrophys. 336 (1998), L45.
\bibitem{Izmodenov_2001} Izmodenov, V., Gruntman, M., Malama, Y., J. Geophys. Res. 106 (2001), 10681.
\bibitem{mica00} Myasnikov,  Izmodenov, V., Alexashov, D., Chalov, S., J. Geophys. Res. 105 (2000), 5179.
\bibitem{maic00} Myasnikov, Alexashov, D., Izmodenov, V., Chalov, S., J. Geophys. Res. 105 (2000), 5167.
\bibitem{Aleksashov_2000} Aleksashov, D., Baranov, V., Barsky, E., Myasnikov, A., Astronomy Letters 26 (2000), 743-749.
\bibitem{Zaitsev_Izmodenov_2001}Zaitsev, N., Izmodenov V., in The Outer Heliosphere: The Next Frontiers, Edited by K. Scherer,
H. Fichtner, H. Fahr, and E. Marsch, COSPAR Colloquiua Series, 11.
Amsterdam: Pergamon Press (2001), 65-69.
\bibitem{isenberg86} Isenberg, P., J. Geophys. Res.  91 (1986), 9965.
\bibitem{holzer72} Holzer, J. Geophys. Res. 77 (1972), 5407.
\bibitem{Lallement_Bertin} Lallement R., Bertin, Astron. Astrophys. 266 (1992), 479-485.
\bibitem{Linsky_1993} Linsky, J., Brown A., Gayley, K., et al., Astron. J. 402 (1993), 694-709.
\bibitem{Lallement_1995} Lallement, R., Ferlet, A., et al., Astron. Astrophys. 304 (1995), 461-474.
\bibitem{Jaeger_Fahr_1998} Jaeger, Fahr, H., Solar Physics, 178 (1998),631-656.
\bibitem{fahr_86} Fahr, H., Neutsch, W., Grzedzielski, S., et al., Space Sci. Rev 43 (1986) 329-381.
\bibitem{Grzedzielski_81}Grzedzielski, S., Macek, W., Obrec, P., Nature, 292 (1981) 615-616.
\bibitem{Grzedzielski_88}Grzedzielski, S., Macek, W., J. Geophys. Res. 93 (1988), 1795-1808.
\end{thebibliography}
\end{document}